\documentclass[10pt]{scrartcl}

\usepackage{jheppub}

\usepackage{booktabs}

\usepackage{amsfonts}

\newcommand{\I}{{i\mkern1mu}}

\newcommand{\Real}{\Re\hspace{-1pt}\mathfrak{e}}

\newcommand{\HiggsLong}{\cite{Hollik:2014bua}}

\newcommand{\MC}{M_{H^\pm}}

\newcommand{\gsim}{\;\raisebox{-.3em}{$\stackrel{\displaystyle >}{\sim}$}\;}

\begin{document}


\thispagestyle{empty}

\def\thefootnote{\fnsymbol{footnote}}

\begin{flushright}
\parbox{2.3cm}{MPP--2015--11\\
DESY--15--017}
\end{flushright}

\vspace{2cm}

\begin{center}

{\large\sc {\bf Two-loop top-Yukawa-coupling corrections to the}}

\vspace{0.4cm}

{\large\sc {\bf charged Higgs-boson mass in the MSSM}}

\vspace{1cm}

Wolfgang Hollik$^1$\footnote{email: hollik@mpp.mpg.de}
and
Sebastian Pa{\ss}ehr$^2$\footnote{email: sebastian.passehr@desy.de}

\vspace*{.7cm}

{\sl
$^1$Max-Planck-Institut f\"ur Physik \\
(Werner-Heisenberg-Institut), \\
F\"ohringer Ring 6, 
D--80805 M\"unchen, Germany
}
\\
\medskip
{\sl
$^2$Deutsches Elektronen-Synchrotron DESY,\\
Notkestra{\ss}e 85, D--22607 Hamburg, Germany
}

\end{center}

\vspace*{2cm}

\begin{abstract}
{}
The top-Yukawa-coupling enhanced two-loop corrections to the charged
Higgs-boson mass in the real MSSM are presented. The contributing two-loop
self-energies are calculated in the Feynman-diagrammatic approach in
the gaugeless limit with vanishing external momentum and bottom mass,
within a renormalization scheme comprising 
on-shell and $\overline{\text{DR}}$ conditions.
Numerical results illustrate the effect 
of the $\mathcal{O}{\left(\alpha_t^2\right)}$~contributions 
and the importance of the two-loop corrections
to the mass of the charged Higgs bosons.
\end{abstract}

\def\thefootnote{\arabic{footnote}}
\setcounter{page}{0}
\setcounter{footnote}{0}

\newpage

\section{Introduction}

Charged Higgs bosons go along with many extensions of 
the  Standard Model, such as supersymmetric versions
of the Standard Model or general  Two-Higgs-Doublet models.
The neutral Higgs-like particle 
with a mass \mbox{$\simeq 125$~GeV}, discovered by
the ATLAS and CMS experiments~\cite{Aad:2012tfa,Chatrchyan:2012ufa},
behaves within the presently still sizeable experimental uncertainties
like the Higgs boson of the Standard Model 
(see~\cite{ATLAS-Higgs-WWW,CMS-Higgs-WWW} for latest results),
but on the other hand leaves ample room for interpretations
within  extended models with a richer spectrum.
A scenario of particular interest thereby is 
the Minimal Supersymmetric
Standard Model~(MSSM)  with two scalar doublets
accommodating five physical Higgs bosons, at lowest order 
given by the light and heavy $CP$-even~$h$ and~$H$,
the $CP$-odd~$A$, and the charged~$H^\pm$ Higgs bosons. 
The discovery of a charged Higgs boson
would constitute an unambiguous sign of physics beyond the Standard Model,
providing hence a strong motivation for searches for the charged Higgs boson.

Experimental searches for the charged Higgs bosons of the MSSM 
(or a more general Two-Higgs-Doublet Model) 
have been performed at LEP~\cite{ADLOchargedHiggs} 
yielding a robust bound of $\gsim 80\, {\rm GeV}$~\cite{LEPchargedHiggs}.
The Tevatron bounds~\cite{Tevcharged} are meanwhile superseeded by the constraints from
the searches for charged Higgs bosons at the LHC~\cite{LHCcharged}.

The Higgs sector of the MSSM can be parametrized at lowest
order in terms of the gauge couplings $g_1$ and $g_2$, 
the mass $m_A$ of the $CP$-odd Higgs boson, and 
the ratio of the two vacuum expectation values, $\tan\beta \equiv v_2/v_1$;
all other masses and mixing angles are predicted in terms of these quantities.
Higher-order contributions, however, give in general 
substantial corrections to the tree-level relations.

The status of higher-order corrections to the masses and mixing angles
in the neutral Higgs sector is quite advanced. 
A  remarkable amount of work has been done for 
higher-order calculations of the mass spectrum, for real SUSY parameters
\cite{Heinemeyer:1998jw,Heinemeyer:1998np,Heinemeyer:1999be,Heinemeyer:2004xw,Borowka:2014wla,Degrassi:2014pfa,mhiggsFD3l,Zhang:1998bm,Espinosa:2000df,Brignole:2001jy,Casas:1994us,Degrassi:2002fi,Heinemeyer:2004gx,Allanach:2004rh,Martin:2001vx}
as well as for complex 
parameters~\cite{Demir:1999hj,Pilaftsis:1999qt,Carena:2000yi,Heinemeyer:2007aq,Frank:2006yh,Hollik:2014wea,Hollik:2014bua}. 
They are based on full one-loop calculations improved by  higher-order
contributions to the leading terms from the Yukawa sector involving
the large top and bottom Yukawa couplings. 
Quite recently, the $\mathcal{O}{\left(\alpha_{t}^{2}\right)}$ terms
for the complex version of the MSSM were
computed~\cite{Hollik:2014wea,Hollik:2014bua};
they are being implemented into the  program 
{\tt FeynHiggs}~\cite{Heinemeyer:1998yj,Hahn:2010te,Hahn:2009zz}.

Also the mass of the charged Higgs boson is affected by higher-order corrections
when expressed in terms of $m_A$. The status is, however, 
somewhat less advanced as compared to the neutral  Higgs bosons. 
Approximate one-loop corrections were already derived 
in~\cite{mhp1lA,mhp1lB,mhp1lD}.
The first complete one-loop calculation in the Feynman-diagrammatic
approach was done in~\cite{mhp1lE}, and more recently the corrections
were re-evaluated in \cite{markusPhD,Frank:2006yh,Frank:2013hba}.
At the two-loop level, important ingredients for the leading corrections
are the  $\mathcal{O}({\alpha_t \alpha_s})$ and
$\mathcal{O}{\left(\alpha_{t}^{2}\right)}$ contributions to the charged 
$H^\pm$ self-energy. The $\mathcal{O}({\alpha_t \alpha_s})$ part
was obtained in~\cite{Heinemeyer:2007aq} for the complex MSSM,
where it is required for predicting the neutral Higgs-boson spectrum 
in the presence of $CP$-violating mixing of all three neutral 
$CP$ eigenstates with the charged Higgs-boson mass used 
as an independent (on-shell) input parameter instead of $m_A$.
In the $CP$-conserving case, on the other hand, with
$m_A$ conventionally chosen as independent input quantity,
the corresponding self-energy contribution has been exploited  for obtaining
corrections of  $\mathcal{O}(\alpha_t \alpha_s)$  to the mass of the charged
Higgs boson~\cite{Frank:2013hba}.
In an analogous way, the recently calculated $\mathcal{O}({\alpha_t^2})$ part of the
$H^\pm$ self-energy in the complex MSSM~\cite{Hollik:2014wea,Hollik:2014bua},
can now be utilized for the real, $CP$-conserving, case to derive
the $\mathcal{O}({\alpha_t^2})$ corrections to the charged Higgs-boson mass
as well.

In the present paper we combine the new
two-loop terms of $\mathcal{O}({\alpha_t^2})$
with the complete one-loop  and 
$\mathcal{O}(\alpha_t \alpha_s)$ two-loop contributions
to obtain an improved prediction for the mass 
of the charged Higgs boson. 
The results have been implemented into the code {\tt FeynHiggs}.
An overview of the calculation is given in Section 2,  
followed by a numerical evaluation and discussion
of the two-loop corrections in Section 3
and Conclusions in Section 4.

\clearpage

\section{Higgs-boson mass correlations}

\subsection{Tree-level relations}

We consider the Higgs potential of the MSSM with real parameters, at the tree-level given by
\begin{align}\label{eq:Potential}
  \begin{split}
    V_{\rm Higgs} &=  m_1^2\,\mathcal{H}_{1}^{\dagger} \mathcal{H}_{1} + m_2^2\,\mathcal{H}_{2}^{\dagger} \mathcal{H}_{2} 
      + \left(m_{12}^2\, \epsilon_{ab} \mathcal{H}_{1}^a \mathcal{H}_{2}^b + \text{h.\,c.}\right)\\
    &\quad + \frac{1}{8}\left(g_{1}^{2} + g_2^{2}\right)\left(\mathcal{H}_{2}^{\dagger}\mathcal{H}_{2} - \mathcal{H}_{1}^{\dagger}\mathcal{H}_{1}\right)^{2} 
    + \frac{1}{2}g_2^{2}\left(\mathcal{H}_{1}^{\dagger}\mathcal{H}_{2}\right)\left(\mathcal{H}_{2}^{\dagger}\mathcal{H}_{1}\right)\,,
  \end{split}
\end{align}
with the mass parameters $m_1^2, m_2^2, m_{12}^2$, 
and the gauge-coupling constants $g_1, g_2$.
The two scalar Higgs doublets in the real MSSM can be decomposed according to
\begin{alignat}{4}
  \label{eq:Higgsfields}
  \mathcal{H}_{1} &= \begin{pmatrix} v_{1} + \frac{1}{\sqrt{2}}(\phi_{1} - \I \chi_{1})\\ -\phi^{-}_{1}\end{pmatrix},&\quad
  \mathcal{H}_{2} &= \begin{pmatrix} \phi^{+}_{2}\\ v_{2} + \frac{1}{\sqrt{2}}(\phi_{2} + \I \chi_{2})\end{pmatrix} ,
\end{alignat}
with real vacuum expectation values~$v_1$ and~$v_2$. The ratio~$\left.v_2\middle/v_1\right.$ is denoted as~\mbox{$\tan\beta \equiv t_\beta$}.
The mass-eigenstate basis is obtained by the transformations
\begin{align}\label{eq:Higgsmixing}
 \begin{pmatrix} h\\ H \end{pmatrix} &= 
   \begin{pmatrix}
    -s_\alpha & c_\alpha\\ c_\alpha & s_\alpha \end{pmatrix}  
 \begin{pmatrix} \phi_{1}\\ \phi_{2}\end{pmatrix},  &\; \;
 \begin{pmatrix} H^{\pm}\\ G^{\pm} \end{pmatrix}   &= 
   \begin{pmatrix}
    -s_{\beta_c} & c_{\beta_c}\\ c_{\beta_c} & s_{\beta_c} \end{pmatrix} 
   \begin{pmatrix} \phi^{\pm}_{1}\\  \phi^{\pm}_{2}  \end{pmatrix}, \; \;
   \begin{pmatrix} A\\  G \end{pmatrix} 
  = \begin{pmatrix} -s_{\beta_n} & c_{\beta_n} \\ c_{\beta_n} & s_{\beta_n} \end{pmatrix} 
    \begin{pmatrix} \chi_{1}\\ \chi_{2}    \end{pmatrix},
\end{align}
[with \mbox{$s_x \equiv \sin{x}$} and~\mbox{$c_x \equiv \cos{x}$} ],
where $h,H,A$ and $H^\pm$ denote the physical neutral and 
charged Higgs bosons, and $G^0, G^\pm$ the unphysical neutral and charged
(would-be) Goldstone bosons.

The Higgs potential in the real MSSM can be written as the following
expansion in terms of the components $h,H,A,H^\pm,G^\pm$
[with $(H^-)^\dagger = H^+$,  $(G^-)^\dagger = G^+$],
\begin{align}\label{eq:HiggsPotential}
  \begin{split}
    V_{\rm Higgs} &= -T_h\,h - T_H\,H  \,
            +\, \begin{pmatrix} H^-, & G^-\end{pmatrix} 
           \begin{pmatrix} m_{H^{\pm}}^2  & m_{H^-G^+}^2 \\ m_{G^-H^+}^2 & m_{G^{\pm}}^2\end{pmatrix} 
           \begin{pmatrix} H^+\\ G^+\end{pmatrix}\\[0.3cm]
          &\quad + \frac{1}{2}\begin{pmatrix} h, & H, & A, & G \end{pmatrix}
            \begin{pmatrix} m_h^2  & m_{hH}^2 & 0 & 0 \\
                                      m_{hH}^2 & m_H^2  &  0  & 0 \\ 
                                       0           &  0         & m_A^2  & m_{AG}^2 \\ 
                                       0           &   0        & m_{AG}^2 & m_G^2 
           \end{pmatrix}
            \begin{pmatrix} h\\ H\\ A\\ G\end{pmatrix} + \dots  \, ,
  \end{split}
\end{align}
omitting higher powers in the field components. 
Explicit expressions for the entries in the mass matrices are given in
Ref.~\cite{Frank:2006yh} for the general complex MSSM [the special case here 
is obtained for setting $T_A =0$ in those expressions].
Of particular interest for the correlation between the neutral
$CP$-odd and the charged Higgs-boson masses are the entries for 
$m_A^2$ and $m_ {H^\pm}^2$, reading
\begin{align}
\label{eq:mamc}
\begin{split}
\begin{alignedat}{5}
&m_A^2     &&= m_1^2\, s_{\beta_n}^2 &&+ m_2^2\, c_{\beta_n}^2 &&+ m_{12}^2\, s_{2\beta_n} 
              &&- \tfrac{1}{4} (g_1^2+g_2^2)(v_1^2 - v_2^2)\,c_{2\beta_n} \, , \\
&m_{H^\pm}^2 &&= m_1^2\, s_{\beta_c}^2 &&+ m_2^2\, c_{\beta_c}^2 &&+ m_{12}^2\, s_{2\beta_c}
              &&- \tfrac{1}{4} (g_1^2+g_2^2)(v_1^2 - v_2^2)\,c_{2\beta_c}
               +\tfrac{1}{2} g_2^2 (v_1 c_{\beta_c} +v_2 s_{\beta_c})^2 \, . 
\end{alignedat}
\end{split}
\end{align}
At lowest order, after applying the minimization conditions for the Higgs potential,
the tadpole coefficients~\mbox{$T_h, T_H$} vanish
and the mass matrices become diagonal
for $\beta_c =\beta_n =\beta$, yielding 
\begin{align}
   m_{H^\pm}^2 &= m_A^2 + M_W^2\,,
 \label{eq:Higgscharged}\\
 m_{h,\,H}^2 &= \frac{1}{2}\left(m_{A}^{2} +
    M_{Z}^{2}\mp\sqrt{\left(m_{A}^{2} + M_{Z}^{2}\right)^{2} - 4 m_{A}^{2} M_{Z}^{2}\, c_{2\beta}^{2}}\right) ,
\end{align}
when $\alpha$ is chosen according to 
\begin{align}
  \label{eq:alpha}
  \tan (2\alpha) &= \frac{m_{A}^2 + m_Z^2}{m_A^2 - m_Z^2}\,  \tan(2\beta)\, , 
 \quad {\rm with} \quad -\frac{\pi}{2} < \alpha < 0 \, .
\end{align}
The Goldstone bosons~$G^0$ and~$G^\pm$ remain massless.

In the following we focus on the modification of the relation~(\ref{eq:Higgscharged})
by higher-order contributions, 
which allows to derive the charged Higgs-boson mass
in terms of the $A$-boson mass $m_A$ and the model parameters
entering through quantum loops.

\subsection{The charged Higgs-boson  mass beyond lowest order}

Beyond the lowest order, the entries of the mass matrix of the charged Higgs bosons 
are shifted by adding their corresponding renormalized self-energies. 
The higher-order corrected mass~$\MC$ of the physical charged Higgs bosons,
the pole mass,  is obtained from the zero of the renormalized two-point vertex function,
\begin{align}
\label{eq:gammacharged}
   \MC^2  = \Real (s_0) \, ,  \qquad
  \left.\hat{\Gamma}_{H^+ H^-}{\left(p^2\right)}\right|_{p^2\,=\,s_0}  &= 
     \I \left[p^2 - m_{H^\pm}^2 + \hat{\Sigma}_{H^+ H^-}{\left(p^2\right)}\right]_{p^2\,=\,s_0} = 0\,  .
 \end{align}
Therein, $\hat{\Sigma}_{H^+ H^-}{\left(p^2\right)}$ denotes the
renormalized self-energy for the charged Higgs bosons~$H^\pm$, 
which we treat as a perturbative expansion,
\begin{align}\label{eq:sigmacharged}
  \hat{\Sigma}_{H^+ H^-}{\left(p^2\right)} &= \hat{\Sigma}_{H^+H^-}^{(1)}{\left(p^2\right)} +
  \hat{\Sigma}_{H^+H^-}^{(2)}{\left(p^2\right)} +\, \cdots \, .
\end{align}
At each loop order $k$, the renormalized self-energy~$\hat{\Sigma}_{H^+ H^-}^{(k)}$ 
is composed of the unrenormalized self-energy~$\Sigma_{H^+ H^-}^{(k)}$ and 
a corresponding counterterm~$\delta^{(k)}m_{H^{\pm}}^{\mathbf{Z}}$, according to
\begin{align}\label{eq:masscorr}
  \hat{\Sigma}_{H^+ H^-}^{(k)}{\left(p^2\right)} &= \Sigma_{H^+ H^-}^{(k)}{\left(p^2\right)} - \delta^{(k)}m_{H^{\pm}}^{\mathbf{Z}}{\left(p^2\right)}\,.
\end{align}
At the one-loop level the counterterm is given by
\begin{align}
\label{eq:chargedSECTone}
  \delta^{(1)}m_{H^{\pm}}^{\mathbf{Z}}{\left(p^2\right)} &=  
  \left(m_{H^{\pm}}^2 - p^2\right)\delta^{(1)}Z_{H^{\pm}H^{\pm}} + \delta^{(1)}m_{H^{\pm}}^2\,,
\end{align}
and at the two-loop level by 
\begin{align}
\label{eq:chargedSECTtwo} 
  \delta^{(2)}m_{H^{\pm}}^{\mathbf{Z}}{\left(p^2\right)} &=
  \left(m_{H^{\pm}}^2\!- p^2\right) \left[ \delta^{(2)}Z_{H^{\pm}H^{\pm}} + \tfrac{1}{4}\left(\delta^{(1)}Z_{H^{\pm}H^{\pm}}\right)^2\right] 
   - p^2 \tfrac{1}{4} \left(\delta^{(1)}Z_{H^\pm G^\pm}\right)^2\\
  &\quad+ \delta^{(1)}Z_{H^{\pm}H^{\pm}}\,\delta^{(1)}m_{H^{\pm}}^2
  +\tfrac{1}{2}\,\delta^{(1)}Z_{H^\pm G^\pm}
  \left(\delta^{(1)}m_{H^-G^+}^2\!+ \delta^{(1)}m_{G^-H^+}^2\right) + \delta^{(2)}m_{H^{\pm}}^2\,
 \nonumber
\end{align}
involving field-renormalization constants and genuine mass counterterms 
of one- and two-loop order; they are specified in Ref.~\HiggsLong, from where
conventions and notations have been taken over and simplified to the case of
the real MSSM.

\begin{figure}[tb]
  \centering
  \includegraphics[width=.82\textwidth]{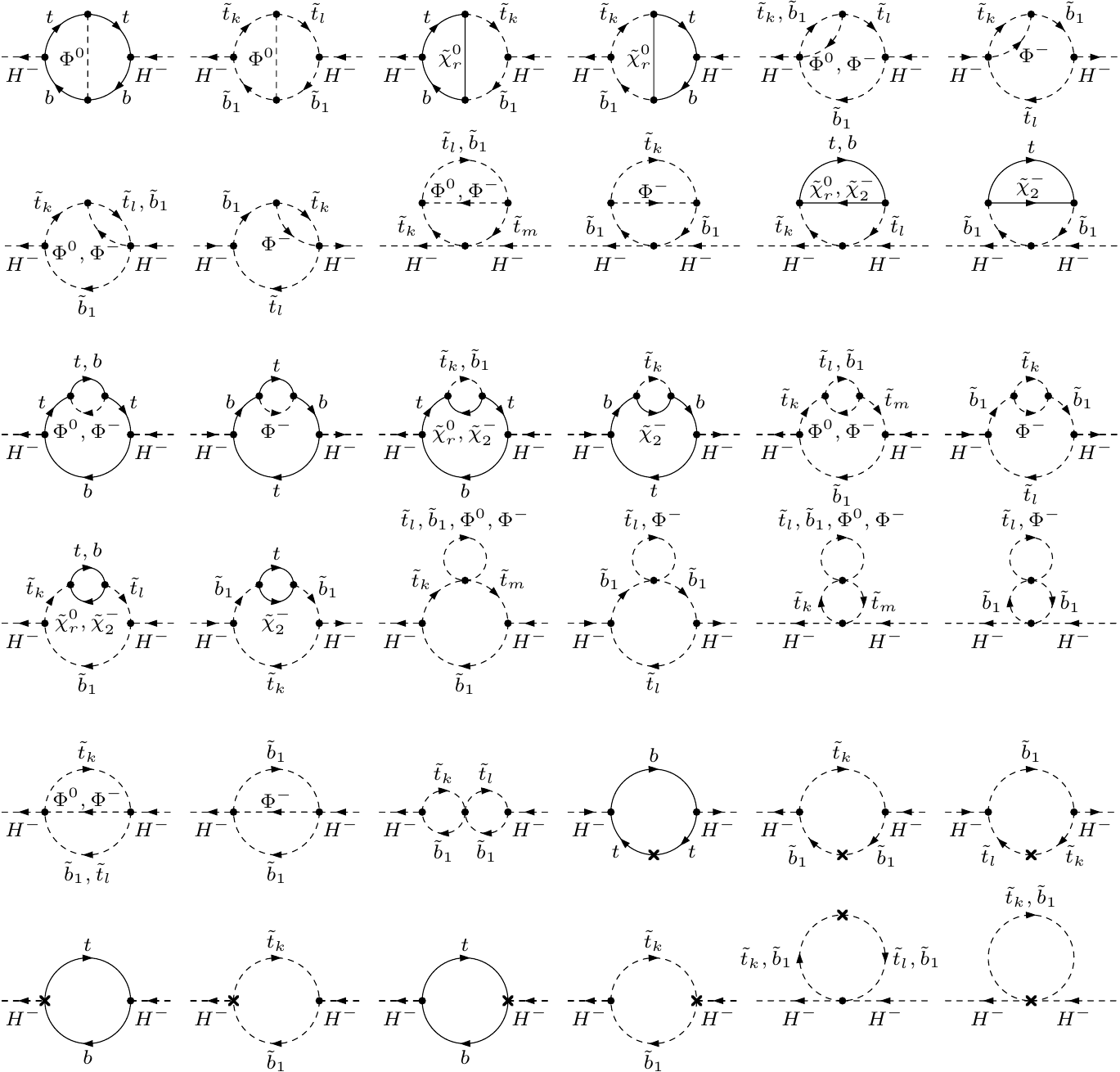}
  \caption[Full list of two-loop self-energy diagrams for the charged
  Higgs bosons]
 {\label{fig:selfenergiescharged}Full list of two-loop self-energy diagrams for the charged Higgs bosons.
  Each cross denotes a one-loop counterterm insertion.
 \mbox{$\;\Phi^0 = h,\,H,\,A,\,G$};  \mbox{$\;\Phi^- = H^-,\,G^-$}.  }

\vspace*{0.5cm}
\end{figure}

Whereas the one-loop self-energy $\hat{\Sigma}_{H^+ H^-}^{(1)} (p^2)$
of the charged Higgs boson is completely known, 
at the two-loop level only results in the approximation for $p^2=0$
have become available, namely the $\mathcal{O}{\left(\alpha_t\alpha_s\right)}$  
corrections calculated earlier~\cite{Heinemeyer:2007aq,Frank:2013hba},
and the two-loop Yukawa contributions~$\mathcal{O}{\left(\alpha_t^2\right)}$ 
which are presented in this paper.
The evaluation of these terms is performed in the gaugeless limit 
and the bottom-quark mass set to zero (as done in
Ref.~\cite{Frank:2013hba}), thus yielding the top-Yukawa-coupling enhanced parts. 
Detailed analytical results of the
two-loop self-energy and renormalization were published in Ref.~\HiggsLong. 
The diagrammatic calculation of the self-energies and counterterms
was performed
with {\tt FeynArts}~\cite{Hahn:2000kx}, {\tt FormCalc}~\cite{Hahn:1998yk},
and {\tt TwoCalc}~\cite{Weiglein:1993hd}.
The full list of Feynman diagrams
of~$\mathcal{O}{\left(\alpha_t^2\right)}$ for the self-energy of the
charged Higgs boson is illustrated in Fig.~\ref{fig:selfenergiescharged}.

\begin{figure}[t]
  \centering
  \includegraphics[width=.82\textwidth]{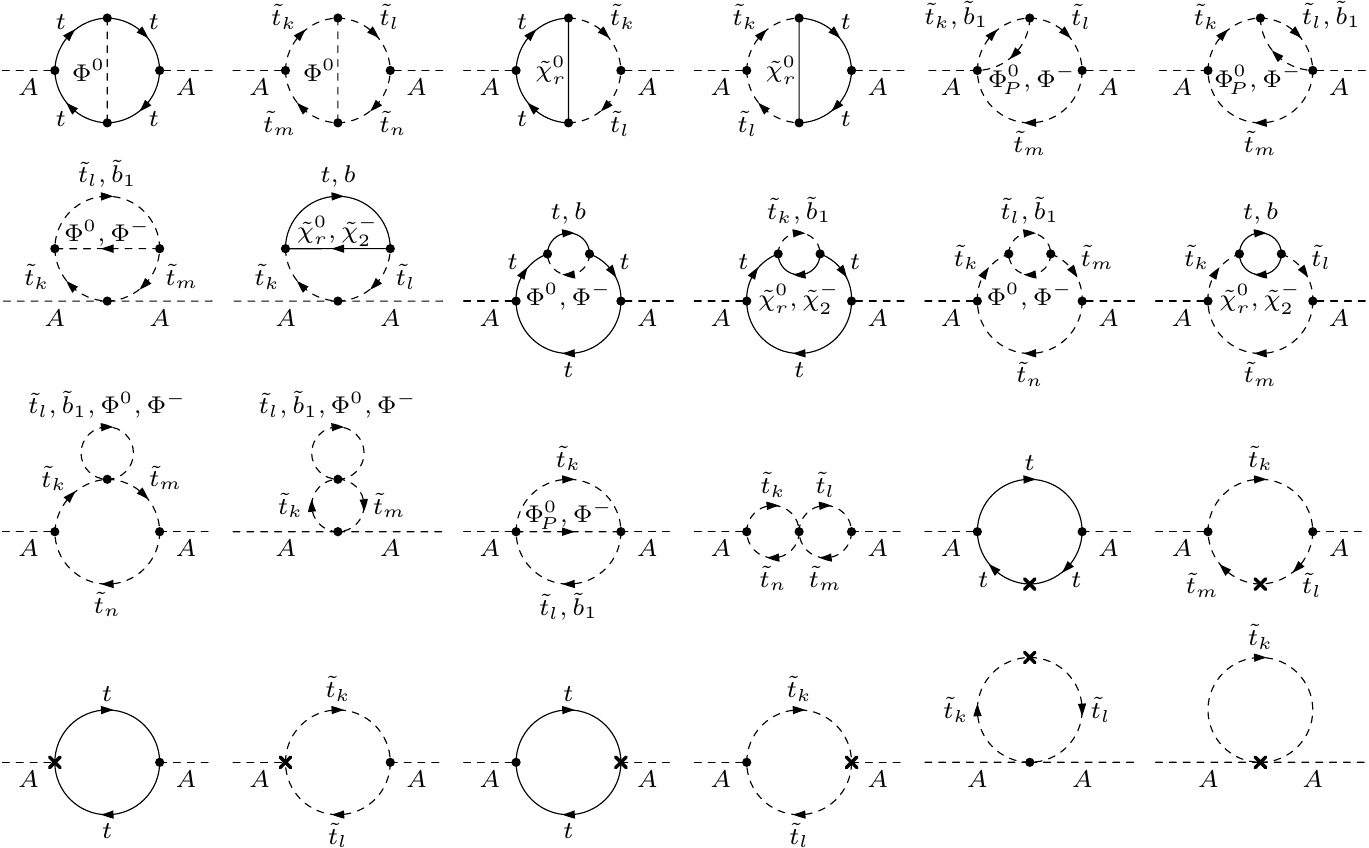}
  \caption[Full list of two-loop self-energy diagrams for the $A$-boson]{\label{fig:selfenergiesA}Full list of two-loop self-energy diagrams for the $A$-boson.
   Each cross denotes a one-loop counterterm insertion.
 \mbox{$\;\Phi^0 = h,\,H,\,A,\,G$}; \mbox{$\;\Phi^0_{\! P} = A,\,G$};  \mbox{$\;\Phi^- = H^-,\,G^-$}.  }

\vspace*{0.5cm}
\end{figure}

Within our approximations for the two-loop part of the charged Higgs-boson self-energy,
\begin{align}
\hat{\Sigma}_{H^+ H^-}^{(2)} (0) & = \Sigma_{H^+ H^-}^{(2)} (0)  -
 \delta^{(2)} m_{H^{\pm}}^{\mathbf{Z}} (0) \, , 
\end{align}
the two-loop counterterm~\eqref{eq:chargedSECTtwo} simplifes to
\begin{align}
\label{eq:chargedCT}
  \begin{split}
  \delta^{(2)}m_{H^{\pm}}^{\mathbf{Z}}{\left(0\right)} &= 
   m_{H^\pm}^2 \left[\delta^{(2)}Z_{H^{\pm}H^{\pm}} + \tfrac{1}{4}\left(\delta^{(1)}Z_{H^{\pm}H^{\pm}}\right)^2\right] + \delta^{(2)}m_{H^{\pm}}^2\\
  &\quad+ \delta^{(1)}Z_{H^{\pm}H^{\pm}}\,\delta^{(1)}m_{H^{\pm}}^2 + 
    \tfrac{1}{2}\,\delta^{(1)}Z_{H^\pm G^\pm}\left(\delta^{(1)}m_{H^-G^+}^2\!+ \delta^{(1)}m_{G^-H^+}^2\right)\,.
  \end{split}
\end{align}
The genuine mass counterterms~$\delta^{(k)}m_{H^{\pm}}^2$ are
determined by Eq.~\eqref{eq:mamc} and setting $\beta_n =\beta_c =\beta$
(see also Ref.~\HiggsLong). In the gaugeless limit they are given by
(for $k=1,2$)
\begin{align}
\label{eq:chargedmassCT}
  \delta^{(k)}m_{H^{\pm}}^2 &= \delta^{(k)}m_A^2  \, .
\end{align}
The other genuine mass counterterms are determined by the relation
\begin{align}
  \delta^{(1)}m_{H^-G^+}^2 = \delta^{(1)}m_{G^-H^+}^2 &= 
  -\frac{e}{2\, s_{\text{w}}\, M_W} \, \delta^{(1)}T_H\, 
      - m_{H^{\pm}}^2\, c_{\beta}^2\; \delta^{(1)}t_{\beta}\, , 
\end{align}
involving the tadpole counterterm $\delta^{(1)}T_{H}$ and the counterterm $\delta^{(1)}t_{\beta}$
for the renormalization of $\tan\beta$.

In the real MSSM,
the mass of the~$CP$-odd Higgs boson~$m_A$ is conventionally chosen as
a free input parameter; it can thus be renormalized on-shell at each order. 
Accordingly, the corresponding renormalization conditions in our present approximation
read in terms of the renormalized $A$-boson self-energy as follows,
\begin{align}\label{eq:mAOS}
  \hat{\Sigma}_A^{(k)}{\left(0\right)}  & =
  \Sigma_A^{(k)}{\left(0\right)} - \delta^{(k)}m_A^{\mathbf{Z}}{\left(0\right)}\, = 0 \, . 
\end{align}
The unrenormalized self-energy ~$\Sigma_A^{(2)}$ corresponds to
the Feynman diagrams depicted in Fig.~\ref{fig:selfenergiesA}.
The counterterms in \eqref{eq:mAOS}
at the one-loop and two-loop level read as follows,
\begin{subequations}
  \begin{align}
    \delta^{(1)}m_A^{\mathbf{Z}}{\left(0\right)} &= m_A^2\,\delta^{(1)}Z_{AA} + \delta^{(1)}m_A^2\,,\\
    \begin{split}
      \delta^{(2)}m_A^{\mathbf{Z}}{\left(0\right)} &= m_A^2 \left[\delta^{(2)}Z_{AA} + \tfrac{1}{4}\left(\delta^{(1)}Z_{AA}\right)^2\right] + \delta^{(2)}m_A^2\\
      &\quad+ \delta^{(1)}Z_{AA}\,\delta^{(1)}m_A^2 + \delta^{(1)}Z_{AG}\,\delta^{(1)}m_{AG}^2\,.
    \end{split}
  \end{align}
\end{subequations}
The one-loop non-diagonal mass counterterm~$\delta^{(1)}m_{AG}^2$ therein is given  by
\begin{align}
  \delta^{(1)}m_{AG}^2 &= - \frac{e}{2\, s_{\text{w}}\, M_W}\, \delta^{(1)}T_H \,
                                      -  m_A^2\, c_{\beta}^2 \;  \delta^{(1)}t_{\beta}\,  . 
\end{align}
From the conditions~\eqref{eq:mAOS} for $k=1,2$ the renormalization constants~$\delta^{(k)}m_A^2$ 
are determined and thus the mass counterterms $\delta^{(k)} m_{H^+}^2$  
for the charged Higgs bosons in Eq.~\eqref{eq:chargedmassCT},
required for the two-loop counterm~\eqref{eq:chargedCT}
in the charged Higgs-boson self energy.
All field-renormalization constants 
$\delta^{(k)}\! Z_{\{AA,AG,H^\pm H^\pm,H^\pm G^\pm\} }$ are linear combinations
of the basic field-renormalization constants $\delta^{(k)}\! Z_{\mathcal{H}_{i}}$ 
for the two scalar doublets~\eqref{eq:Higgsfields}, 
as given in Ref.~\HiggsLong.

In addition to the mass counterterms $\delta^{(k)} m_A^2$, the
independent renormalization constants required 
for renormalization of the charged Higgs-boson self-energy are:
the field renormalization constants~$\delta^{(1)}Z_{\mathcal{H}_{i}}$, 
the renormalization constant~$\delta^{(1)}t_{\beta}$ for $\tan\beta$, and
the tadpole renormalization constants~$\delta^{(1)}T_h$, $\delta^{(1)}T_H$
(the two-loop field renormalization constants  
cancel in the renormalized self-energies in the $p^2=0$ approximation).
Moreover, for the one-loop subrenormalization, we need the counterterms 
for the top quark and squark masses
$\delta^{(1)}m_t$, $\delta^{(1)}m_{\tilde{t}_1}$, $\delta^{(1)}m_{\tilde{t}_2}$
and for the trilinear coupling $\delta^{(1)} \! A_t$, 
as well as  the counterterm for the bilinear coefficient of the superpotential, $\delta^{(1)} \! \mu$. 
They are fixed in the same way as described in Ref.~\HiggsLong\
and we do not repeat them here.

\section{Numerical analysis}

In this section we compute numerically the charged Higgs-boson mass~$\MC$ in the real~MSSM in terms of~$m_A$ 
chosen as an input parameter. 
For this purpose, we combine in  the renormalized charged Higgs-boson self-energy 
our new $\mathcal{O}{\left(\alpha_t^2\right)}$~contribution
described in the previous section with the already known
complete one-loop term and the $\mathcal{O}{\left(\alpha_t\alpha_s\right)}$~contribution,
\begin{align}
\label{eq:sigchargedall}
\hat{\Sigma}_{H^+H^-} (p^2) & = \hat{\Sigma}_{H^+H^-}^{(1)} (p^2) 
                                              + \hat{\Sigma}_{H^+H^-}^{(\alpha_t\alpha_s)} (0)
                                              + \hat{\Sigma}_{H^+H^-}^{(\alpha_t^2)} (0) \, ,
\end{align}
as the currently best approximation for~\eqref{eq:sigmacharged}. 
The resulting charged Higgs-boson mass~$\MC$ is obtained via Eq.~\eqref{eq:gammacharged} 
with the help of {\tt FeynHiggs}.

In the following numerical analysis we use  the input parameters as listed in
Tab.~\ref{tab:parameters} (giving also those parameters not needed for
the two-loop self-energies, but required for specifiying the input for 
the other terms in~(\ref{eq:sigchargedall})  and for {\tt FeynHiggs}).
The other parameters of the~MSSM not contained in Table~\ref{tab:parameters} 
are kept variable and are given in the figures.
The quantities~$\mu,\, t_{\beta}$ and the Higgs field-renormalization constants are defined 
in the~$\overline{\text{DR}}$ scheme at the scale~$m_{t}$ 
(see also Ref.~\HiggsLong\ for more details).

\begin{table}[h]
  \centering
  \caption[Default input parameters]{Default input values of the MSSM and SM parameters.}
  \label{tab:parameters}
  \begin{tabular}{r@{$\;=\;$}lr@{$\;=\;$}l}
    \toprule
    \multicolumn{2}{c}{MSSM input} & \multicolumn{2}{c}{SM input}\\
    \midrule
    $M_2$ & $200$~GeV, & $m_t$ & $173.2$~GeV,\\
    $M_1$ & $\left.\left(5s_{\rm w}^2\right)\middle/\left(3c_{\rm w}^2\right) M_2\right.$, & $m_b$ & $4.2$~GeV,\\
    $m_{\tilde{l}_1} = m_{\tilde{e}_{\rm R}}$ & $2000$~GeV, & $m_{\tau}$ & $1.777$~GeV,\\
    $m_{\tilde{q}_1} = m_{\tilde{u}_{\rm R}} = m_{\tilde{d}_{\rm R}}$ & $2000$~GeV, & $M_W$ & $80.385$~GeV,\\
    $A_u = A_d = A_e$ & $0$  & $M_Z$ & $91.1876$~GeV,\\
    $m_{\tilde{l}_2} = m_{\tilde{\mu}_{\rm R}}$ & $2000$~GeV, & $G_{\text{F}}$ & $1.16639\cdot 10^{-5}$,\\
    $m_{\tilde{q}_2} = m_{\tilde{c}_{\rm R}} = m_{\tilde{s}_{\rm R}}$ & $2000$~GeV, & $\alpha_s$ & $0.118$,\\
    $A_c = A_s = A_\mu$ & $0$.\\
    \bottomrule
  \end{tabular}
\end{table}

\smallskip
The influence of the~$\mathcal{O}{\left(\alpha_{t}^{2}\right)}$ corrections on the charged Higgs-boson mass 
decreases with increasing values of $t_\beta$, where the top Yukawa coupling is diminished. 
Therefore we constrain our analysis on values of~$t_\beta < 10$. In
the case of larger~$t_\beta$ also the corrections of~$\mathcal{O}{\left(\alpha_{b}\alpha_{t}\right)}$ may become
relevant (see also Ref.~\cite{Frank:2013hba} for more discussions on the validity range). 

\begin{figure}[tb]
  \centering
  \includegraphics[width=\textwidth]{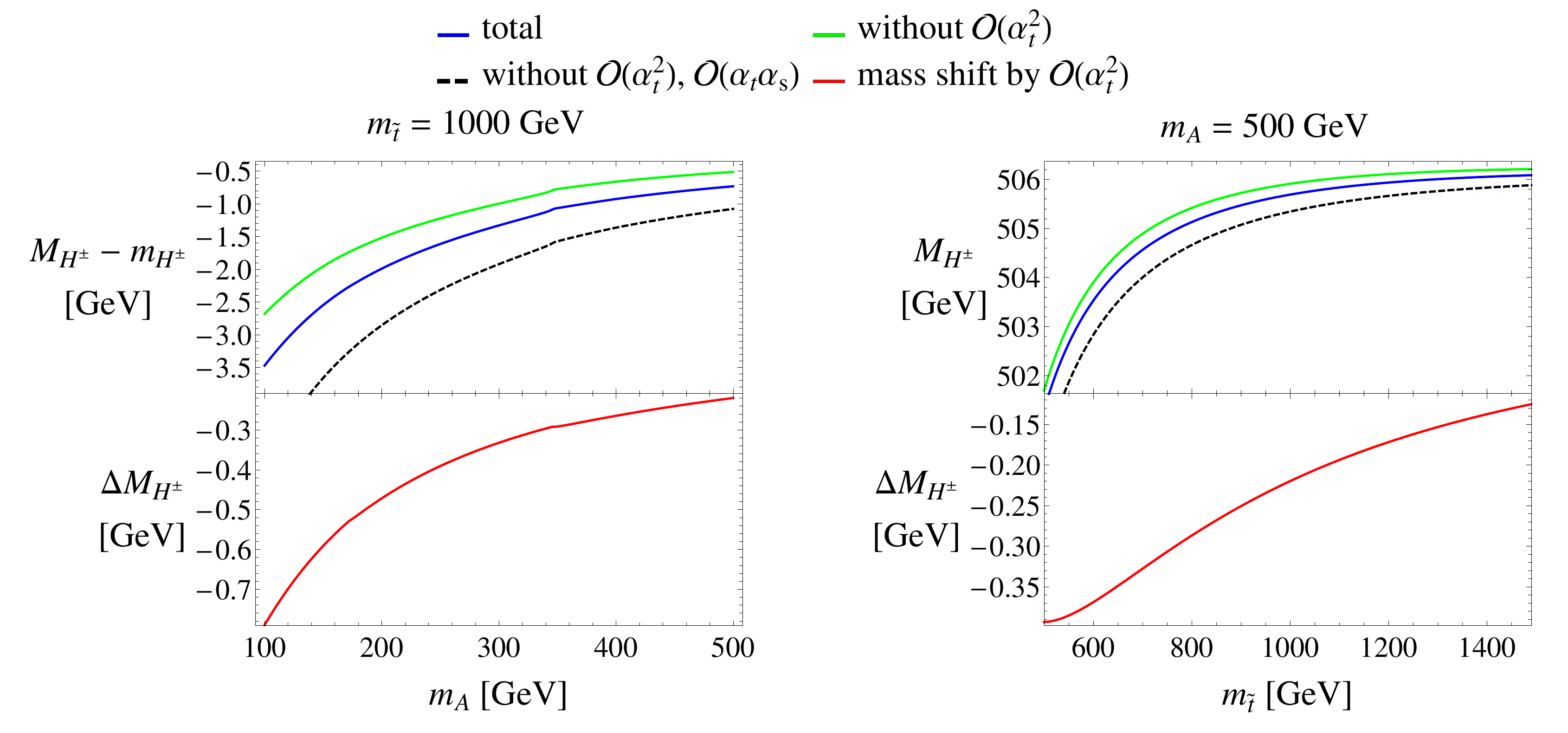}
  \caption{\label{fig:HpmAmSt}%
    Upper parts: prediction for the charged Higgs-boson mass~$\MC$
    including all known contributions~(blue), without
    the~$\mathcal{O}{\left(\alpha_{t}^{2}\right)}$~contributions~(green)
    and without any two-loop corrections~(black dashed) 
    [$m_{H^\pm}$ is the tree-level mass according to Eq.~\eqref{eq:Higgscharged}].
    Lower parts: the mass shift~$\Delta \MC$ by the~$\mathcal{O}{\left(\alpha_{t}^{2}\right)}$~contributions (red).
   Left:~\mbox{$m_{\tilde{t}} \equiv m_{\tilde{q}_{3}} =
    m_{\tilde{t}_{\text{R}}} = m_{\tilde{b}_{\text{R}}} =
    1000$~GeV}. Right:~\mbox{$m_A = 500$~GeV}. The other input
  parameters are \mbox{$t_{\beta} = 8$}, \mbox{$\mu = 2000$~GeV},
  \mbox{$m_{\tilde{\ell}_3} = m_{\tilde{\tau}_{\rm R}} = 1000$~GeV},
  \mbox{$X_t = 2\,m_{\tilde{t}}$}, \mbox{$A_b = A_{\tau} = 0$},
  \mbox{$m_{\tilde{g}} = 1500$~GeV}, for both cases.}
\end{figure}

The shifts in the charged Higgs-boson mass resulting from
the~$\mathcal{O}{\left(\alpha_{t}^{2}\right)}$~contributions are in
general small. In Fig.~\ref{fig:HpmAmSt} the dependence of ~$\MC$  
on the Higgs-sector input parameter~$m_A$ and on the
third-generation soft-breaking squark mass parameter
\mbox{$m_{\tilde{t}} \equiv m_{\tilde{q}_3} = m_{\tilde{t}_{\text{L}}} = m_{\tilde{t}_{\text{R}}}$}
is depicted, showing a decreasing size of the two-loop  mass shift~(red) for increasing
values of both variables. 
The upper section of the figure shows the charged
Higgs-boson mass as obtained at the one-loop level (dashed),
and  with the inclusion of 
the~$\mathcal{O}{\left(\alpha_{t}\alpha_{s}\right)}$~contributions~(green)
and also the~$\mathcal{O}{\left(\alpha_{t}^2\right)}$~terms~(blue). 
The lower section of Fig.~\ref{fig:HpmAmSt}
shows the mass shift originating solely from 
the~$\mathcal{O}{\left(\alpha_{t}^2\right)}$ two-loop part.
Thereby,
the~$\mathcal{O}{\left(\alpha_{t}^2\right)}$~corrections appear as
negative, thus diminishing the two-loop contribution 
of~$\mathcal{O}{\left(\alpha_{t}\alpha_{s}\right)}$. In total, the
two-loop terms still 
yield a positive shift upon the one-loop result for~$\MC$.

\begin{figure}[tb]
  \centering
  \includegraphics[width=\textwidth]{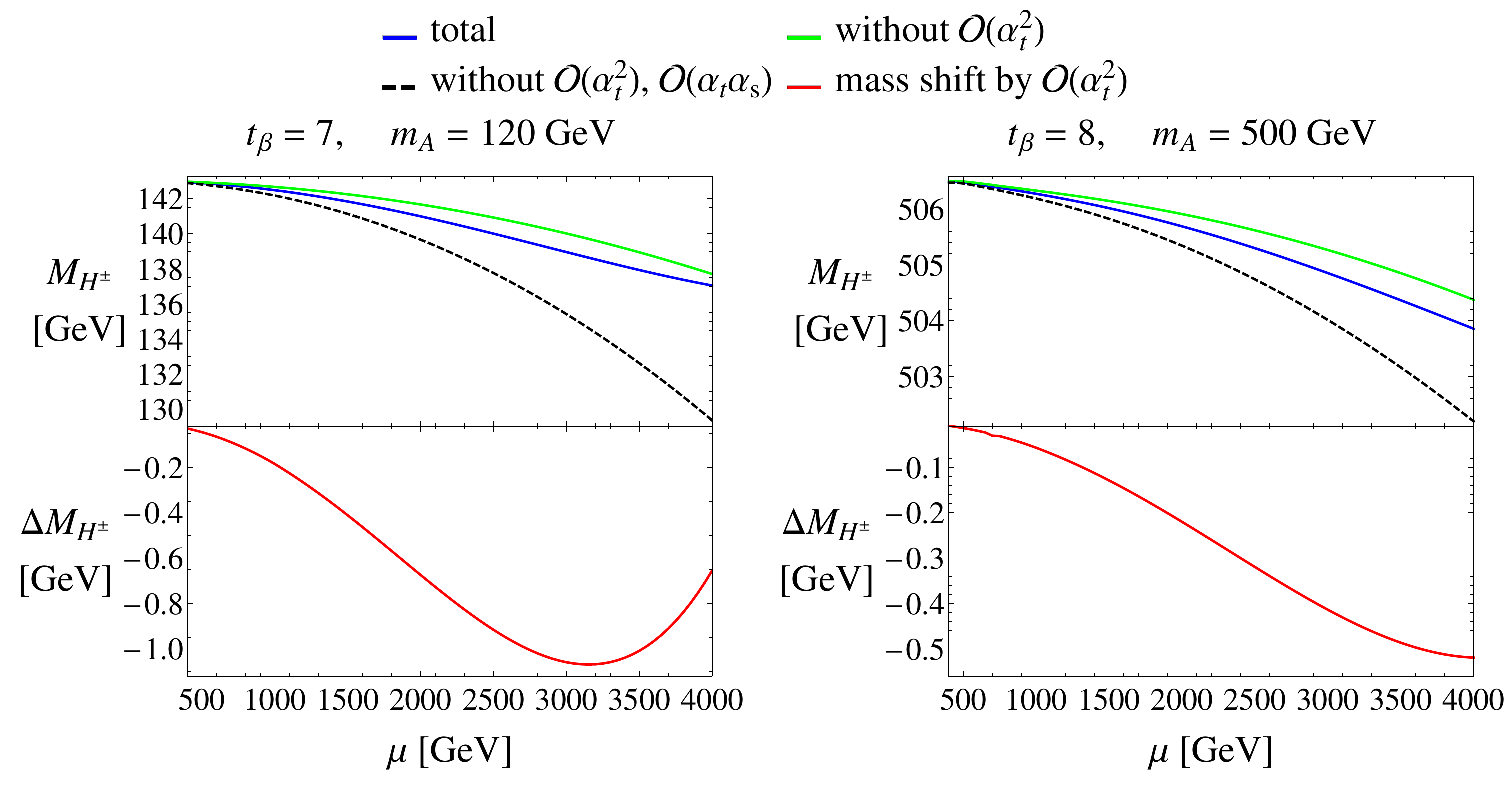}
  \caption[Mass prediction and mass shift for the charged Higgs boson]{\label{fig:Hpmureal}%
    Upper parts: prediction for the charged Higgs-boson mass~$\MC$ including all known contributions~(blue), without the~$\mathcal{O}{\left(\alpha_{t}^{2}\right)}$~contributions~(green) and without any two-loop corrections~(black dashed). Lower parts: the mass shift~$\Delta \MC$ by the~$\mathcal{O}{\left(\alpha_{t}^{2}\right)}$~contributions (red).
  Left:~\mbox{$t_{\beta} = 7$}, \mbox{$m_A = 120$~GeV}, \mbox{$A_t = 2.5\,m_{\tilde{q}_{3}}$}, \mbox{$A_b = A_{\tau} = 0$}.
  Right:~\mbox{$t_{\beta} = 8$}, \mbox{$m_A = 500$~GeV}, \mbox{$X_t = 2\,m_{\tilde{q}_{3}}$}, \mbox{$A_b = A_{\tau} = 0$}.
  The other input parameters are \mbox{$m_{\tilde{q}_{3}} = m_{\tilde{t}_{\text{R}}} = m_{\tilde{b}_{\text{R}}} = 1000$~GeV},
  \mbox{$m_{\tilde{\ell}_3} = m_{\tilde{\tau}_{\rm R}} = 1000$~GeV}, \mbox{$m_{\tilde{g}} = 1500$~GeV}, for both cases.}
\end{figure}

Fig.~\ref{fig:Hpmureal} contains the charged Higgs-boson mass $\MC$,
together with the two-loop shift of~$\mathcal{O}{\left(\alpha_{t}^2\right)}$, 
for a typical~low-$m_H$ scenario (left)~\cite{Carena:2013qia} 
and for a scenario with heavier $H^\pm$ (right),
versus the Higgsino mass~$\mu$.
For large values of~$\mu$, the charged Higgs-boson 
mass~$\MC$ decreases, but the mass shift~$\Delta\MC$ resulting from
the~$\mathcal{O}{\left(\alpha_{t}^{2}\right)}$~contributions becomes more sizeable,
reaching 1~GeV and more for the low~$\MC$ case.
In the scenario shown in the right panel of Fig.~\ref{fig:Hpmureal} the two-loop contributions
are smaller in comparison to the one in the left panel, which is a consequence of the smaller
Yukawa couplings for larger values of $m_A$ and $\tan\!\beta$.

\clearpage

In all cases, the~$\mathcal{O}{\left(\alpha_{t}^{2}\right)}$
contributions appear with negative sign and reduce slightly the
positive mass shift arising from $\mathcal{O}{\left(\alpha_{t}\alpha_{s}\right)}$.
In general, the combined two-loop corrections 
result in a positive shift,  which can amount to several GeV,
on top of the one-loop prediction for $\MC$.

\begin{figure}[t]
  \centering
  \includegraphics[width=0.6\textwidth]{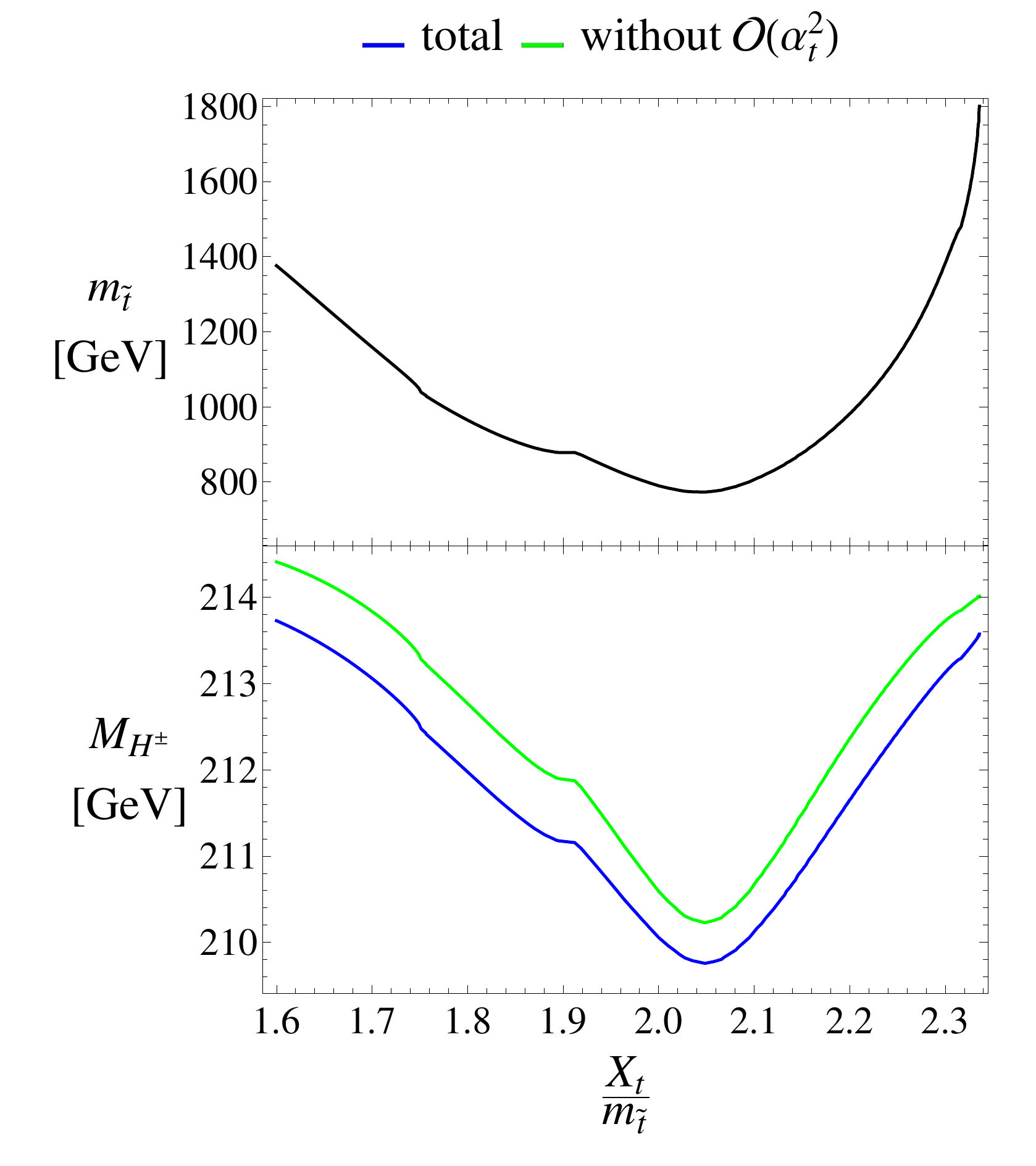}
  \caption{\label{fig:higgsboundsimproved}%
    Mass $M_{H^\pm}$ of the charged Higgs boson with all available
    higher-order terms and without the
    $\mathcal{O}{\left(\alpha_{t}^{2}\right)}$ contributions, for
    ranges of $m_{\tilde{t}}$ and $X_t$ allowed by {\tt HiggsBounds}
    and $m_h=125$~GeV. Other parameters are \mbox{$m_A=
      200$~GeV}, \mbox{$t_\beta = 8$}, \mbox{$\mu = 3000$~GeV},
    \mbox{$A_b = A_{\tau} = 0$}, \mbox{$m_{\tilde{\ell}_3} =
      m_{\tilde{\tau}_{\rm R}} = 1000$~GeV}, \mbox{$m_{\tilde{g}} =
      1500$~GeV}.}
\end{figure}

In the figures mentioned above, the constraint $m_h =125\pm 1\, {\rm GeV}$ on the
light Higgs-boson mass is imposed, except for the low-$m_H$
scenario in Fig.~\ref{fig:Hpmureal} (left) where it is the heavier $H$-boson 
that appears with a mass around $125$~GeV
(a scenario which may soon be excluded by more
stringent limits on the charged Higgs-boson mass).
One has to keep in mind, however, that not all of the parameter values
in the figures, which are shown for illustrating  the parameter
dependence, will actually be allowed when more comprehensive phenomenological studies
on the properties of the Higgs particle at $125$~GeV will be performed.
We have added such a more comprehensive analysis by probing the regions
compatible with the experimental constraints by means 
of  the program {\tt HiggsBounds}~\cite{Bechtle:2008jh,Bechtle:2011sb,Bechtle:2013wla}. The result is shown 
in Fig.~\ref{fig:higgsboundsimproved}, where the $\mathcal{O}{\left(\alpha_{t}^{2}\right)}$
effects for $M_{H^\pm}$ are displayed for possible combinations of the
stop-sector parameters. Also here we find negative mass shifts in the
typical range from $-0.5$~GeV to $-0.8$~GeV.

\section{Conclusions}

We have calculated the two-loop $\mathcal{O}{\left(\alpha_{t}^{2}\right)}$ 
contributions to the mass $\MC$ of the charged Higgs boson
when derived from the $A$-boson mass $m_A$ as an on-shell 
input parameter within the real, $CP$-conserving, MSSM 
and combined them with the complete one-loop 
and the two-loop $\mathcal{O}{\left(\alpha_{t}^{2}\right)}$ contributions.
We have presented numerical studies for scenarios of current
phenomenological interest and discussed the effects of the various
two-loop terms.  

The $\mathcal{O}{\left(\alpha_{t}^{2}\right)}$ two-loop corrections
appear with opposite sign and smaller size with respect to the
$\mathcal{O}{\left(\alpha_{t}\alpha_{s}\right)}$  contributions; 
in combination, the two-loop terms yield
a positive shift to the mass of the charged Higgs boson 
as calculated at one-loop order. 
This shift in $\MC$ can be at the level of several GeV 
and thus of a size that may be relevant
for the LHC (and a future electron-positron collider).

The set of two-loop corrections considered here are expected to be
particularly relevant in parameter ranges of the real MSSM where the 
top-Yukawa terms provide a good approximation to the complete one-loop
result, especially for relatively low values of $\tan\!\beta$  and $m_A$.
In this range, besides precise mass predictions,
the experimental constraints on the mass and the   phenomenological features 
of the lightest Higgs are important and play a substantial role 
when comprehensive analyses withing the MSSM Higgs sector are performed.

Our results for the charged Higgs-boson mass have become part of the Fortran
code {\tt FeynHiggs}.

\section*{Acknowledgement}
This work has been supported by the Collaborative Research Center
SFB676 of the DFG, "Particles, Strings and the early Universe".


\end{document}